\documentclass{caps}
\usepackage{peter}
\usepackage{graphicx}

\def\lessim{\lower 0.5ex\hbox{${}\buildrel<\over\sim{}$}}
\def\gtrsim{\lower 0.5ex\hbox{${}\buildrel>\over\sim{}$}}

\begin{document}

\author{Robert C. Duncan\\\it University of Texas at Austin \ TX \ USA}

\chapter{Triggers of magnetar outbursts}

\abstract{
Bright outbursts from Soft Gamma Repeaters (SGRs)  and Anomalous X-ray Pulsars 
(AXPs) are believed to be caused by instabilities in ultramagnetized neutron 
stars, powered by a decaying magnetic field. 
It was originally thought that 
these outbursts were due to 
reconnection instabilities in the magnetosphere, reached via slow 
evolution of magnetic footpoints anchored in the crust. 
Later models considered sudden shifts in the crust's 
structure.  Recent observations of magnetars give 
evidence that at least some outburst episodes involve rearrangements 
and/or energy releases within the star. 
We suggest that bursting episodes in magnetars are episodes of 
rapid plastic 
yielding in the crust, which trigger ``swarms" of reconnection 
instabilities in the magnetosphere. Magnetic energy always dominates; 
elastic energy released within the crust does not generate strong enough 
Alfv\'en waves to power outbursts. 
We discuss the physics of SGR giant flares, 
and describe recent observations which give 
useful constraints and clues.}

\section{Introduction: A neutron star's crust}

The crust of a neutron star has several components: (1) a Fermi sea of 
relativistic electrons, which provides most of the pressure in the outer 
layers; (2) another Fermi sea of neutrons in a pairing-superfluid
state, present only at depths
below the ``neutron drip" level where the mass-density exceeds 
\hbox{$\rho_{\rm drip}\approx 4.6 \times10^{11}$ gm cm$^{-3}$;} and (3) 
an array of positively-charged nuclei, arranged in a solid (but probably not
regular crystalline) lattice-like structure 
throughout much of the crust. These nuclei become heavier 
and more neutron-bloated at increasing depths beneath the surface, until 
the swollen nuclei nearly ``touch" and the quasi-body-centered-cubic 
nuclear array dissolves 
into rod-like and slab-like structures near the base of the crust: 
``nuclear spaghetti and lasagna" or ``nuclear pasta" (e.g., Pethick \& 
Ravenhall 1995).

In a magnetar, the crust is subject to strong, evolving magnetic 
stresses. 
Magnetic evolution within the crust occurs via Hall 
drift; while ambipolar diffusion and Hall drift of magnetic flux within 
the liquid 
interior strains the crust from below (Goldreich \& Reisenegger 1992; 
Thompson \& Duncan 1996, hereafter ``TD96"). 
The crust and the magnetic field thus 
evolve together through a sequence of equilibrium states in which 
magnetic stresses are balanced by material restoring forces. Because a
magnetar's field is so strong, this  
evolution inevitably involves episodes of crust failure driven by the magnetic 
field on a variety of time-scales. Many complexities are likely to 
affect the crustal yielding threshold, and cause it to vary from place 
to place within a neutron star's crust. Moreover the very nature of 
neutron star crust failure is somewhat uncertain. 
In the deep crusts of 
magnetars, yields may resemble sudden and sporadic flow in an inhomogeneous
liquid crystal, induced by evolving magnetic stresses.

Before discussing these complications, it is worthwhile to note how 
neutron star crust material, outside the pasta layers, differs from
a terrestrial solid. A key difference (due to the high mass-density in a
neutron star) is  
that the relativistic Fermi sea of electrons is only 
slightly perturbed by the Coulomb forces of the nuclei, and does not 
efficiently screen nuclear charges. 
With a nearly inert and uniform 
distribution of negative charge, pure neutron star crust-matter 
comes close to realizing the ``ideal Coulomb crystal."  
Such a body-centered cubic (bcc) lattice is expected to 
form in the 
low-temperature limit of a ``one-component plasma" 
(e.g., Brush, Sahlin 
\& Teller 1966; Ichimaru 1982; van Horn 1991).
The crystallization temperature is 
\hbox{$k_{\scriptscriptstyle B} T_c = \Gamma^{-1} (Ze)^2/a$,} where 
$Z$ is the ionic charge, $a$ is the Wigner-Seitz radius satisfying 
\hbox{${4\over3}\pi a^3 = n^{-1}$} with ion density $n$, and $\Gamma$ 
is a numerical constant found in statistical mechanics to be 
$\Gamma\approx 170$. 
In the deep crust of a neutron star, $T_c \sim 
10^{10}$ K.

The electrostatic structure of naturally-occurring terrestrial solids is 
more complex, with bound electrons and 
efficient screening. Only in the cores of old white 
dwarfs, which have 
cooled sufficiently to crystallize, does bulk material like the stuff of 
a neutron star's outer crust ($\rho < \rho_{drip}$) exist elsewhere in 
nature. Inner-crust matter is found nowhere outside of neutron stars.

However, nearly-ideal Coulomb crystals have recently been made in the 
laboratory, and the failure of these crystals under stress was studied  
(Mitchell et al 2001). I will now describe 
one of these delicate and elegant experiments. 

About 15,000 cold $^9$Be$^+$ ions were confined within a
volume about half a millimeter in diameter, in the laboratory at the
National Institute of Standards in Boulder, Colorado. 
A uniform magnetic field
confined the ions radially, while a static electric field with a 
quadratic potential trapped them axially (i.e., a Penning trap). The 
ions were cooled to millikelvin temperatures using lasers that were 
tuned to a frequency just below an ionic ground-state excitation
level. The plasma crystallized into a disk, with a $bcc$ lattice 
structure. 
(Note that trapping fields effectively provided the 
neutralizing background for this crystal.) Due to a weak radial 
component of the electric field, the charged crystal experienced \bf $E 
\times B$ \rm drift and rotated. The velocity of rotation was controlled 
and stabilized by the experimenters using a perturbing electric field. 
The ions, fluorescing in laser light and separated by 15 microns, were 
directly imaged and photographed. The crystal was then stressed by 
illuminating it with laser light from the side, and off-axis. Slips in 
the crystalline structure were detected. They were distributed, over at 
least three orders of magnitude, according to a power law with index 
between 1.8 and 1.2. 

One can't avoid mentioning here that this distribution of 
slip sizes resembles the 
energy-distribution of bursts from SGRs. Some would 
claim that this is  
a consequence of ``universality" in self-organized critical systems.
In any case it can have no deep implications about
burst mechanisms, since the 
physics of SGR outbursts is much more complex than simple slips in a 
crystal.

These experiments verify some expectations about idealized, crystalline
 neutron star 
matter; but real neutron star
crusts, and especially magnetar crusts, are likely to be complex and messy. 
As emphasized by Ruderman (1991), the circumference of a neutron star is 
about $10^{17}$ lattice-spacings, which is similar to the size of the 
Earth measured in the lattice-spacings of terrestrial rock.  

According to the old, 20th century neutron star theory, the amount 
and distribution of crystalline imperfections in the crust is 
history-dependent. Two factors were thought to be involved: the 
rapidity of cooling 
when the solid originally formed, and the subsequent ``working" of the 
solid by stresses and (in some cases) episodic reheating.
More rapid initial cooling and solidification would generally produce more
lattice imperfections and smaller lattice domains (i.e., smaller grains). 
Extremely rapid 
cooling, or ``quenching" would produce an amorphous (glassy) solid rather than 
a crystal, which is really a long-lived metastable state: a super-cooled
liquid. It was suggested that this occurs in neutron star crusts 
(Ichimaru et al.~1983); however, models of crust solidification in 
more realistic, neutrino-cooled neutron stars showed crystallization (de 
Blasio 1995). 
Subsequent strain-working of the crust would 
increase lattice imperfections, while episodes of 
(magnetic) re-heating followed by cooling 
would tend to anneal the solid.

That's the old picture. Jones (1999, 2001) recently turned  
the story on its head.  He showed that, when the crust initially cools 
through the melting temperature, a substantial range of Z 
(i.e., nuclear proton-numbers) get frozen-in 
at every depth below neutron drip.  This is due to thermal 
fluctuations
in the nuclei, which are in equilibrium with the neutron bath. The energy
separation 
of magic-number proton shells in the neutron-bloated nuclei is not 
large compared to the melt
temperature.   This is important because it means that the crust of
a 21st-century neutron star is amorphous
rather than crystalline.  There exists some short-range 
crystalline order, but
over distances greater than about $\sim 10 \, a$ the variable Z's affect the 
inter-nuclear spacing enough to destroy all order.  This has important 
implications for transport properties such as electrical conductivity, among
other things. 

Finally, there is the sticky issue of nuclear pasta (Pethick \& Ravenhall 
1995 and references therein).\footnote{Nuclear 
pasta results from the competition between 
Coulomb and nuclear surface-energy terms when minimizing the energy. 
The pasta ground states exhibit spontaneously broken symmetries, although 
the underlying interactions between constituent nucleons are nearly 
rotationally-symmetric. This is different from the case with 
terrestrial liquid crystals, in which highly-anisotropic inter-atomic 
forces give rise to the 
broken global symmetries.} 
Deep inside the crust, the 
neutron-bloated nuclei become elongated and join into ``nuclear 
spaghetti": long cylindrical structures 
in a 2-D triangular array. As depth and density 
increase, these nuclear noodles join into 
slabs: ``nuclear lasagna." At even higher 
densities this gives way (at least for some values of nuclear state 
parameters) to ``inverse spaghetti": an array of cylindrical holes in the 
high-density fluid; followed by ``inverse meatballs":
a bcc lattice of spherical holes in otherwise continuous nuclear matter. 
Beneath that lies continuous nuclear fluid.

Because rod-like (or planar) structures can freely slide past each other 
along their length (and breadth) without affecting the Coulomb energy, 
nuclear pasta has an extremely anisotropic tensor of elasticity. Indeed, 
the elastic response of nuclear spaghetti resembles that of the {\it 
columnar phases of a liquid crystal,} and elastic nuclear lasagna resembles the 
{\it smectics A phase} (Pethick \& Potekhin 1998). At sufficiently high
temperatures, positionally-disordered ({\it nematic}) phases are also 
possible; but the threshold for this is $\sim 10^{10}$ K.

It has been suggested that up to half the mass of a neutron star's crust 
is in nuclear pasta 
(Pethick \& Potekhin 1998). A detailed, realistic understanding of the 
response of a neutron star's crust to evolving stresses (especially stresses
largely exerted from below by a magnetic field, as likely in a magnetar) may 
require understanding the size and coherence of nuclear pasta domains; 
their orientation relative to the vertical; 
the interactions of pasta with the magnetic field; and the 
yielding behavior of such 
liquid crystals, which plausibly depends upon instabilities in the
pasta domain structure.
 These difficult issues have not begun to be addressed 
by astrophysicists. 

Thus the range of complicating factors which could affect neutron star 
crust evolution is formidable. In the case of a magnetar, the crust is
coupled to an evolving ultra-strong magnetic field and its generating
currents, which penetrate the underlying core as well. Manifold uncertainties 
about 
field geometries and magnetic evolution compound the 
murkiness. 
Observations of SGRs and AXPs could 
provide the most sensitive probes of neutron star interiors available to 
astronomers, because magnetars are much less stable than other
neutron stars. However, the intertwined complexities of 
neutron star magnetic activity must be unraveled. 

In this review we focus on one aspect of this problem:  
the triggering of bright outbursts.
We will try to keep the discussion on a basic physical level, 
eschewing equations as much as possible.  In Section 2
we review the physics and phenomenology of SGR outbursts. In Section 3 
we compare 
rise-time observations with models for trigger mechanisms. Section 4 
discusses other observations which offer clues. 
Section 5 discusses the general issue of crust-failure in magnetars.
Section 6 gives 
conclusions.

\section{Magnetar outbursts: a brief review}

\smallskip

\it

``Flares are triggered in magnetically-active main-sequence stars when 
convective motions displace the footpoints of the field sufficiently to 
create tangential discontinuities, which undergo catastrophic 
reconnection. Similar reconnection events probably occur in magnetars, 
where the footpoint motions are driven by a variety of diffusive 
processes." \ \ \ \ - Duncan \& Thompson (1992)

\rm

\smallskip

This quotation shows that magnetar outbursts were originally conceived 
as being triggered in the magnetosphere, as a consequence of the neutron 
star's slow, interior magnetic evolution. This was believed to 
apply to giant flares as well: ``The field of a magnetar carries 
sufficient energy to power the 1979 March 5th event ($5\times10^{44}$ 
ergs at the distance of the LMC, assuming isotropic emission; Mazets et 
al.~1979)." (DT92)
Note that the March 5th event was the only giant flare 
which had been detected at that time, with energy $> 200$ times greater 
than the second most-energetic SGR event. 

Paczy\'nski (1992), in work done soon after DT92, made this point more
explicitly.  He suggested that the March 5th event was ``caused by a 
strong magnetic flare at low optical depth, which led to a thermalized
fireball..."
 
By 1995, Thompson and I had realized that the evolving, strong 
field of a magnetar was capable of straining the crust more severely 
than it could bear. Thompson \& Duncan 1995 thus discussed the 
relative merits of impulsive crustal shifts 
and pure magnetospheric 
instabilities as outburst triggers. TD95 favored scenarios in which both 
processes occurred, with the 1979 March 5th event involving profound 
exterior reconnection. 
Thompson \& Duncan 1995 and 
1996 also discussed plastic deformation of magnetar 
crusts. Besides high plasticity at places where the temperature 
approaches $\sim 0.1 \, T_c$ (plausibly due to local magnetic 
heating), we noted that magnetic stresses dominate 
elastic stresses if \hbox{$B > B_\mu = (4\pi \mu)^{1/2} = 4\times 
10^{15} \, \rho_{14}^{0.4}$ G,} 
where $\mu$ is the shear modulus, and $\rho_{14}$ is the mass-density
in units of $10^{14}$ gm cm$^{-3}$.
A magnetic field stronger than $B_\mu$ 
is thus like a 600-pound gorilla: ``it does whatever it wants" in the 
crust. We noted possible implications of plastic deformation for 
glitches, X-ray light curve variations, and 
triggering catastrophic reconnection 
(TD95; TD96; Thompson et al.~2000; TD01).

Observations made after 1996 have tended to fill in the ``energy gap" 
between the 1979 March 5th event and other SGR outbursts. In particular, 
the 1998 August 27th giant flare was about $10^{-1}$ times as energetic 
as the March 5th event (Hurley et al.~1999a; Mazets et al.~1999a; Feroci
et al.~1999); and two intermediate-energy events\footnote{Here, I call 
intermediate-energy events {\it flares} but not {\it giant flares}. Events
releasing $\sim 10^{41}$ ergs or less are traditionally called {\it bursts}.
I use {\it outburst} as a generic term for all magnetar events.}
 have been observed: the 2001 April 18 
flare from SGR 1900+14 (Kouveliotou et al.~2001; Guidorzi et al.~2003) 
and the slow-rising 1998 June 18 
flare from SGR 1627-41 (Mazets et al.~1999b) which had no 
long-duration, oscillating soft tail. 
The emerging continuity in outburst energies 
makes it more plausible that giant flares and 
common SGR bursts differ in degree rather than in kind; while the 
profound differences between outbursts of comparable energies 
indicate that a wide variety of physical conditions and processes are 
involved.

Studies of SGR burst statistics since 1996 have also yielded important 
insights. Cheng, Epstein, Guyer \& Young (1996) noted that the 
statistical distribution of SGR burst energies is a power law with index 
1.6, resembling the Guttenberg-Richter law for earthquakes. Such a 
distribution can result from self-organized criticality 
(Katz 1986; Chen, Bak \& Obukhov 1991).
Cheng et al.~found additional 
statistical resemblances between SGR bursts and earthquakes, as verified and 
further studied by 
G\"o\u{g}\"u\c{s} et 
al.~(1999 and 2000) with a much larger sample of SGR events. 
AXP 2259+586's June 2002 active episode showed very similar burst statistics 
(Gavriil, Kaspi \& Woods 2003). These results 
lend support to the hypothesis that SGR/AXP outbursts are powered by 
an intrinsic stellar energy source, which is plausibly magnetic. 
(Accretion-induced events, including Type I and Type II X-ray bursts, 
have much different statistics.) However, as G\"o\u{g}\"u\c{s} et 
al.~pointed out, these burst statistics do not necessarily argue that 
SGR bursts are 
crustquakes. Similar statistical distributions have been found in solar 
flares (Crosby, Aschwanden \& Dennis 1993; Lu et al.~1993). 

In 2001, Thompson and I studied an idealized ``toy model" for a giant flare. 
In this model, a circular patch of crust facilitates the release   
of magnetic energy by yielding along circular fault 
and twisting.\footnote{The common, 0.1 s SGR bursts, on the other hand, ``could
be driven by a more localized and plastic deformation of the crust." (TD01)}
Circular crust displacements are 
plausible because the crust is stably-stratified and strongly
constrained in its motion, yet significant twisting movement could be 
driven by the magnetic field. Moreover, a magnetar's formation as a 
rapid rotator should significantly ``wind up" the star's interior field 
(DT92); and twists of the exterior field, which would result from this 
kind of magnetic activity, could drive currents through the 
magnetosphere, contributing to the observed, quiescent X-ray emissions 
from magnetars (Thompson et al.~2000).

Note that the crust-yielding event in a giant flare, if it occurs, is not a 
brittle fracture.  Neutron star crusts probably undergo plastic
failure, at least outside the nuclear pasta (Jones 2003).   
The sudden yielding event could have been a widely-distributed plastic 
flow along
circular flow-lines induced by rapidly-changing stresses
exerted by the core field from below.

Alternatively, there may exist instabilities within the crust that 
cause rapid mechanical 
failure, triggering giant flares (and/or 
other events).
The failure of nuclear pasta could involve sudden instabilities due to 
the interactions of domains with differently-oriented, strongly anisotropic
elastic/liquid response.  In the solid crust, 
a sufficiently long and localized plastic slip 
could drive melting along the fault, suppressing the normal elastic 
stress and mimicking a brittle fracture. Jones (2003) estimates that 
this could occur for slips longer than a few centimeters. The process 
might be facilitated, as a ``mock-fracture" propagates, by the development 
of magnetic gradients 
within the fault plane, with localized magnetic heating. 

In 2002, Thompson, Lyutikov \& Kulkarni (hereafter ``TLK") considered the 
possibility that giant flares are instabilities which develop in the 
magnetosphere with 
no energetically-significant crust displacement on the 
time-scale of the flare. 
Section 5.6 of TLK discussed four pieces of observational evidence which 
bear on the question of which mechanism operates. TLK argued that three 
out of four favored crust-yielding. Finally, Lyutikov (2003) 
gave a new estimate of the rise-time for 
magnetospheric instabilities in magnetars. 
Since rise times 
are an important diagnostic we now discuss them in detail.

\section{Outburst rise-times and durations}

There are two rise-times of interest in SGR outbursts: the ``growth time" 
$\tau_{grow}$ which is the e-folding time for the energy-flux growth 
during the initial, rapid brightening; and the ``peak time" $\tau_{peak}$ 
which is the time from the initial onset of the event until the 
(highest) peak of the light curve. 

A third time-scale of interest in the brightest SGR events is 
$\tau_{spike}$, the duration of the initial, hard-spectrum, extremely 
bright phase of the event which we refer to as the ``hard spike." In 
both giant flares on record, this spike is followed by 
an intense ``soft tail" of X-rays, modulated on the rotation 
period of the star. 
The soft tail is thought to be emitted by an optically-thick ``trapped 
fireball" in the magnetosphere of a magnetar (TD95). The abrupt 
vanishing of soft tail emission at the end of the 1998 August 27th event 
seems to be due to fireball evaporation, corroborating this 
interpretation (Feroci et al.~2000; TD01). 

The March 5th 1979 event reached its peak at $\tau_{peak} \approx 20$ 
ms, but the initial, fast rise through many orders of magnitude was 
unresolved by ISEE or the Pioneer Venus
Orbiter, thus $\tau_{grow}<0.2$ ms (Cline et al.~1980, 
Terrel et al.~1980; Cline 1982). The initial, hard-spectrum
emission lasted for $\tau_{spike}\sim 0.15$ s (Mazets et al.~1979), 
during which time it showed variability
on timescales of order $\sim 10 - 30$ ms (Barat et al.~1983).

Thompson and I suggested interpretations of these time scales. TD95 
noted (in eq.~16) that the Alfv\'en crossing time within the 
(fully-relativistic) magnetosphere of a magnetar is comparable to the 
light-crossing time of the star, roughly 30 microseconds. ``Since 
reconnection typically occurs at a fraction of the Alfv\'en velocity, the 
growth time of the instability is estimated to be an order of magnitude
larger$\ldots$  This is, indeed, comparable to the 0.2 msec rise time 
of the March 5 event." In other words, we suggested
\begin{equation}
\tau_{grow} \sim  {L\over 0.1 \, V_A} \sim  0.3 \left({L\over 10 \, \hbox{\rm 
km}}\right)\ \hbox{\rm ms,}
\end{equation}
where $L$ is the scale of the reconnection-unstable zone, and $V_A \sim 
c$ is the (exterior) Alfv\'en velocity. (Observations of solar flares give
evidence that reconnection often proceeds at speeds $\sim 0.1 \, V_A$; 
e.g.~Dere 1996.) 
TD95 further suggested that 
$\tau_{spike}$ is comparable to the {\it interior} Alfv\'en wave crossing time 
of the star, which applies if the event involves an interior 
magnetic rearrangement. This yields \hbox{$\tau_{spike} \sim 0.1 \, 
B_{15}^{-1} \, \rho_{15} \, (\Delta \ell / R_\star)$ s} [TD95 eq.~17], 
in agreement with giant flare data: $\tau_{spike} = 0.15$ s [March 5th
event] and 0.35 s [August 27th event].

Another physical time scale of possible relevance is the shear-wave 
crossing time of the active region of crust. (This is the
elastic stress-equilibration timescale even when the generation of 
propagating shear waves is small.) The shear-wave velocity \hbox{ $V_\mu
= (\mu / \rho)^{1/2}$} is insensitive to depth (or local density 
$\rho$) in the crust, at least in the zones outside the nuclear pasta: 
\hbox{$V_\mu = 1.0 \times 10^{3} \, \rho_{14}^{-0.1}$ km s$^{-1}$} 
[TD01, eq.~8]. For an active region of size $\ell$, 
this gives a crossing time
\hbox {$\tau_\mu = \ell/V_\mu = 3 \, (\ell/ 3$ km$)$ ms} (TD95), 
thus $\tau_{grow}\le \tau_\mu \le \tau_{peak}$ for the March 5th flare.

The light curves of common, repeat bursts from SGRs were studied by 
G\"o\u{g}\"u\c{s} et al.~(2001), using a data-base of more than 
900 bursts from two SGRs that were observed using the Rossi X-ray Timing 
Explorer (RXTE). 
G\"o\u{g}\"u\c{s} et al.~found that 
the distribution of burst durations (as measured by
$T_{90}$, the time in which 90\% of the 
burst counts accumulate)
is lognormal, with a peak of order 100 ms. 
Most bursts rise faster 
than they decline, but many have roughly triangular light curves. 
In particular, about half of all bursts have $\tau_{peak} > 0.3 \, T_{90}$. 
Thus the distribution of 
$\tau_{peak}$ peaks at $\sim 30$ ms, and 
$\tau_{grow}$ peaks around $\sim 10$ ms. 

Lyutikov (2003) studied the growth of spontaneous reconnection in magnetar 
magnetospheres. Compared to better-understood conditions in the Solar 
chromosphere, radiative cyclotron decay times are extremely short, 
forcing currents to flow narrowly along field lines. 
The plasma is thus 
force-free and relativistic, being dominated by the magnetic field. 
Lyutikov suggested that a tearing-mode instability operates within 
current-sheets, involving the clustering of current filaments within the 
sheet and the formation of ``magnetic islands." This 
has a rate \hbox{$\tau_{rise} \sim \sqrt{\tau_A 
\tau_R}$,} just as in the non-relativistic case,
where $\tau_A \sim \ell/c$ is the Alfv\'en crossing time of 
the unstable zone, and $\tau_R$ is the resistive time-scale 
\hbox{$\tau_R \sim \ell^2/\eta$}.
Lyutikov conjectured that either Langmuir turbulence or ion sound 
turbulence provide the resistivity: $\eta \sim c^2/ \omega_p$, 
where $\omega_p$ is the (electron or ion) plasma frequency. To evaluate 
this requires an estimate of the local particle density. 
Lyutikov adopted a value expected in the ``globally-twisted magnetosphere" 
model of TLK, which yields\footnote{Note that the electron plasma 
frequency is approximately \hbox{$\omega_p \sim \sqrt{\omega_B \, c / r} \sim 
3 \times 10^{11}$ rad/s}, where the cyclotron frequency $\omega_B = (eB/mc)$ 
is evaluated at $r =100$ km, outside a $R = 10$ km star with polar field 
$B_{pole} =5 \times 10^{14}$ Gauss, assuming $B(r ) = B_{pole} 
(r/R)^{-2-p}$ with $p= 1/2$ in a strongly twisted magnetosphere, 
$\Delta \phi\approx 2$ radians.} 
\begin{equation}
\tau_{grow} \sim 0.1 \left({L \over 10 \, \hbox{\rm km}}\right)^{3/2} \
\left({r\over 100 \, \hbox{\rm km}}\right)^{-7/8} \
\left({B_{pole} \over 5 \times 10^{14}\, \hbox{\rm G}}\right)^{1/4} \
\hbox{\rm s,}
\end{equation}
for Langmuir turbulence, 
or smaller by a factor $(m_p/m_e)^{1/4}=6.5$ for
ion sound turbulence. If reliable, this analysis represents an 
improvement over TD95's crude estimate [eq.~(1) above]. But to match 
observed rise-times requires quite localized events, high in the
magnetosphere, with fully-developed ion turbulence in the tearing layer. 
If the events happen closer to the stellar surface ($r\sim 10$ km) as 
likely,\footnote{The fraction of 
exterior magnetic energy lying beyond radius $r$, $f_B(>r)$ falls off 
substantially {\it faster} than the pure dipole contribution 
\hbox{$f_B(>r) = (r/R)^{-1-2p}$,} where $p = 1$ for no global twist.  So the
fraction of energy available for reconnection at $r > 100$ km is significantly
less than $10^{-2}$ [$p = 1/2$; 2-radian twist] or $10^{-3}$ [untwisted].}
then the 
rise time is closer to $\sim 1$ s for electron turbulence and $\sim 0.1$ s for 
ions. 
The problem is that this mechanism requires low particle 
densities $n$ to proceed quickly: 
\hbox{$\tau_{grow}\propto \omega_p^{1/2} \propto n^{1/4}$.} 

More realistic models of the magnetosphere than a simple global twist 
will greatly exacerbate the discrepancy, because reconnection 
occurs where the current density $j$ (and thus $n \sim [j/q v]$, where $q$ is 
the charge and usually $v\sim c$) is especially large, in current sheets. That 
is, the magnetosphere can be locally as well as globally twisted; and the
current density is determined by local magnetic shear. 

Other mechanisms 
besides tearing modes coupled to ion sound turbulence probably operate 
in nonrelativistic astrophysical 
reconnection, 
and seem worthy of investigation in the magnetar context. 
One possibility is stochastic reconnection (Lazarian \& Vishniac 1999),
which requires some source of turbulence on scales that are larger 
than the
current sheet width.  In a magnetar this might be provided by crust-yielding
motion which agitates the field near a developing magnetic discontinuity.

\section{Other observational clues}

There is evidence that at least some magnetar outbursts involve 
structural adjustments inside the star, with enhancements of magnetospheric
currents:

$\bullet$ The 1998 June 18 event from SGR 1627-41 resembled a 
slow-rising giant flare with no soft tail (Mazets et al.~1999b). One 
plausible interpretation is that the star experienced a deep stellar
adjustment that triggered little exterior reconnection or other energy 
dissipation in zones of low-lying, closed field-lines
(relative to other powerful flares) and thus no long-lasting trapped
fireball. This is consistent
with the results of Kouveliotou et al.~(2003), who studied X-ray
emissions from SGR 1627-41 following June 1998. For two years, the light 
curve was a 0.47-index power law, gradually leveling off to a 
``plateau";
and then, after 1000 days, dropping precipitously. Kouveliotou et al.~found
that the cooling crust of a $10^{15}$ Gauss neutron star could
follow this pattern if the initial energy deposition (presumably on June 
18th) extended deep into the crust, significantly below neutron drip.
The integrated energy of the X-ray afterglow was comparable to
the June 18 outburst energy; but the impulsive energy injection in the
crust had to be much larger (by a factor $\sim 10^2$ in Kouveliotou et
al.'s models) because of deep conduction and neutrino losses. This would 
then be a (relatively) crust-active, magnetosphere-quiet magnetar. One
concern with this interpretation is that the observed afterglow spectrum 
was non-thermal and time-variable. It has not yet been shown that
reprocessing by scattering in the magnetosphere can (fully) account for
this. Other interpretations of SGR 1627-41 data might still be
possible.

$\bullet$ \ The June 18 event had a total duration $\Delta t \sim 0.5$ 
s, comparable to a magnetar's internal Alfv\'en-crossing time 
(cf.~TD95).  This flare also
peaked much more gradually than other flares: \hbox{$\tau_{peak} \sim
0.1$ s}.
This could be consistent with slow, catastrophic crust
failure, say at a rate $V \sim 0.1 V_\mu$, along a large
fault-line or plastic shear-zone: $\tau_{peak}\sim 0.1 (\ell / 10$ km) s 
(Mazets et al.~1999b). Note that a zone of crust
adjusting over a timescale $\tau \gtrsim 0.1$ s would produce little Alfv\'en
wave emission on field lines shorter than $c \cdot \tau \sim 3\times 10^4$ km.
If the energy of this flare was released mostly on far-reaching
field lines, then it would tend to blow these field lines open where 
the field is weak, far from the star, and/or
promptly radiate from a large emiting zone at limited optical
depth, rather than create a long-lasting, optically-thick, trapped 
photon-pair plasma.
This could explain how such a short-duration event could attain peak 
luminosity
\hbox{$L_{peak} \simeq 3 \times 10^{44} \, D_{11}^2$ erg s$^{-1}$}
with a hard spectrum, comparable to the peak luminosity of a giant flare,
at a distance \hbox{$D = 11 \, D_{11}$ kpc} (Corbel et al.~1999).

$\bullet$ \ Timing studies of AXP 1E2259+586 revealed a glitch 
associated with a burst-active episode in June 2002, plausibly 
simultaneous with the onset of bursting (Kaspi et al. 2003; Woods et al. 
2003b). The star's rotation rate abruptly increased by 4.2 parts in 
$10^6$, giving evidence for the redistribution of angular 
momentum between superfluid and non-superfluid components within the 
star. A sudden adjustment within the 
star is necessary, thus the bursting episode was probably not 
due to pure magnetospheric instabilities.\footnote{Note that 
the June 2002 glitch in AXP 2259+586 (Kaspi et 
al.~2003; Woods et al.~2003) was different from previous spindown 
irregularities in this star during the past 25 years. I say this because
the star has been spinning down at a steady rate during the 
$\sim 5$ years since phase-coherent timing began (Kaspi, Chakrabarty \& 
Steinberger 1999; Gavriil \& Kaspi 2002), and the persistent 
(post-recovery) $\dot{P}$ changed by only $\sim 2\%$ during the 
glitch/bursting episode. If one extrapolates with this $\dot{P}$ back through 
the sparsely-sampled period history of the star, beginning with {\it Einstein} 
Observatory observations in 1979, one finds that the star must have 
experienced two episodes of {\it accelerated spindown,} or two spin{\it 
down} glitches, both with $(\Delta P / P) \sim + 2 \times 10^{-6}$. The 
first occurred around 1985, between Tenma and EXOSAT observations. The 
second occurred after ASCA but before RXTE, during 1993-1996. One could 
alternatively fit the data with spin-{\it up} glitches, like the June 
2002 glitch, as suggested by Usov (1994) and Heyl \& Hernquist (1999), 
but this 
fit requires that the persistent value of $\dot{P}$ was larger in the 
past by $\sim 25\%$. In either case, this star was behaving differently 
in the past.}

$\bullet$ No significant, persistent diminishment of $\dot{P}$ was 
detected in SGR 1900+14 following the 1998 August 27 giant flare (Woods 
et al. 1999). This puts constraints on large-scale rearrangements of the 
magnetosphere (Woods et al.~2001) in the context of the globally-twisted
magnetosphere model (TLK). 
In this model, twists are maintained in the force-free magnetosphere 
by currents flowing along field lines. 
Such twists could be driven by a strong, ``wound up" 
interior field, stressing the crust from below, which is a likely
relic of magnetar formation (DT92). 
As shown by TLK, global twists tend to shift field lines 
away from the star, enhancing the field strength at the light cylinder 
and hence the braking torque. If the August 27 flare involved a 
relief of large-scale twists via reconnection, analogous to 
the instabilities of 
Wolfson (1995) and Lynden-Bell \& Boiley (1994), with significant 
diminishment of global currents (a possibility raised by TLK and 
Lyutikov 2003), then one would expect a diminishment in $\dot{P}$. \ In 
fact, there was no significant change in $\dot{P}$ immediately after the 
flare, but $\dot{P}$ significantly {\it increased} over the years which 
followed (Woods et al.~2002; Woods 2003a; Woods 2003b). This gives 
evidence that the net effect of the 1998 magnetic activity episode, 
including the giant flare, was to {\it increase} the global 
twist angle and global currents, in a way which did not immediately affect the
near-open field lines, far from the star (C. Thompson, private 
communication). A complete discussion of SGR torque variations will be 
given elsewhere. Here I simply want to point out that models of giant
flares which posit that the whole magnetosphere is restructured, 
with largely dissipated currents, are not supported by SGR timing data.

This concludes my short list of new evidence. Thompson, Lyutikov \& Kulkarni 
(2002; \S 5.6) gave three additional semi-empirical arguments for crustal 
shifts 
during the August 27 flare. They also gave one countervailing 
argument, based upon the softening of SGR 1900+14's spectrum after the 
giant flare. But later work (Lyubarski, Eichler \& Thompson 2002) 
suggested that the immediate post-burst 
emission was dominated by surface afterglow with a soft, thermal 
spectrum (which is presumably modified by scattering outside the star).

Lyutikov (2003) noted five pieces of evidence which favor 
magnetospheric instabilities. His first point was based on 
X-ray pulse-profile changes in SGR 1900+14. These same observations
were invoked by TLK to 
argue the other way, so I think that this evidence is ambiguous. (See \S 5
in Woods et al.~2003a for a complete discussion.)
Some of
his other points might have alternative interpretations, as he himself 
notes. In particular, the mild statistical anti-correlation between
burst fluence and hardness (G\"o\u{g}\"u\c{s} et al 2002) could be due 
to emission-physics effects,  
independent of trigger details, as 
suggested by G\"o\u{g}\"u\c{s} et al.  Incidentally, this 
mild trend seems
to go the other way in AXP bursts (Gavriil, Kaspi \& Woods 2003).

\section{Discussion: crust failure in magnetars}

A magnetar's crust is a degenerate, inhomogeneous Coulomb solid 
in a regime of high pressure, 
magnetization, and stress which has no direct experimental analog. It 
lies atop a magnetized liquid crystal which is subject to unbearable
Maxwell stresses from the core below.
Its behavior is thus quite uncertain.
The crust cannot fracture like a brittle terrestrial solid, which develops
a propagating crack with a microscopic void (Jones 2003),
but it may 
experience other instabilities,
such as ``mock fractures" 
(\S 2 above). 

The magnetar model was developed in a series of papers by Thompson and myself
which invoked magnetically-driven crust fractures.  
Most of this work will remain  
valid if magnetar crusts prove to 
yield only plastically with no instabilities. 
However, some of the outburst physics 
would return closer to the original 
conception of 
DT92 and Paczy\'nski (1992), along with several other changes. 

For example, TD96 considered ``Hall fracturing" in the crusts of 
magnetars. This was a consequence of ``Hall drift" 
(Jones 1988; Goldreich \& Reisenegger 1992), whereby the Hall term in 
the induction equation drives helical wrinkles in the magnetic field, 
which stress the crust. In magnetars, but not in radio pulsars, the 
wrinkled field is strong enough to drive frequent, small-scale crust failure 
as ambipolar 
diffusion within the core forces flux across the crust from below. TD96 
suggested that this crust failure occurs in small fractures 
which generate a quasi-steady flux of high-frequency Alph\'en waves, 
energizing the magnetosphere and driving a diffuse wind out from the 
star. There seemed to exist direct evidence for this: a bright, compact 
radio nebula that was believed to surround SGR 1806-20 (Kulkarni et 
al.~1994; Sonobe et al.~1994; Vasisht et al.~1995; Frail et al.~1997).  
However, the high power of this nebula required 
rather implausibly optimized physical parameters
and efficiencies, which caused concern.\footnote{ Because the 
magnetar model predicted that SGR 1806-20 was rotating slowly (as later 
verified; Kouveliotou et al.~1998) this nebula could not have been 
rotation-powered. Fracture-driven Alfv\'en waves from an active, 
vibrating crust thus seemed necessary, 
beginning in October 1993 (when the 
radio nebula discovery was announced at the Huntsville GRB Workshop: Frail \& 
Kulkarni 1994; Murakami et al.~1994) 
through most of the 1990's. The probability for a chance 
overlap of the radio plerion core with the $\sim 1$ arcmin ASCA X-ray 
box for the SGR (Murakhami et al.~1994) was initially estimated in the 
range $\lessim 10^{-6}$. When a coincident, extremely reddened Luminous 
Blue Variable (LBV) star was discovered (Kulkarni et al.~1995; van 
Kerkwijk et al. 1995) smaller probabilities for 
chance coincidence were 
implied, so the LBV star was presumed to be a binary companion to the 
plerion-powering neutron star. It turns out that the LBV star may be
the brightest star in the Galaxy, with
luminosity $L > 5 \times 10^6 \, L_\odot$ 
(Eikenberry et al.~2002; Eikenberry et al.~2003), thus it probably 
can power the radio nebula by itself (e.g.,~Gaensler et al.~2001; Corbel
\& Eikenberry 2003). The chance 
for this $M > 200 \, M_\odot$ star to lie within $14^{\prime\prime}$ of 
another nearly-unique galactic star, SGR 1806-20, is fantastically small. 
This seems to be a lesson in the dangers of 
{\it a posteriori} statistics, or the treachery of Nature, or both.} 
Then Hurley et al.~(1999b) found evidence that SGR 1806-20 
is not precisely coincident with the central peak of the radio nebula. 
{\it Chandra} 
measurements verified that the SGR is 
displaced 14$^{\prime\prime}$ from the radio core 
(Eikenberry et al.~2001; Kaplan et al.~2002). 
It now seems likely that many, or perhaps all, Hall-driven yields in 
a magnetar's crust 
are plastic, occurring via 
dislocation glide in the outer crust ($\rho < \rho_{drip}$) and microscopic
shear-layers at depth.  This probably dissipates 
magnetic 
energy locally as heat, rather than as Alfv\'en waves in a corona.
Still the basic analysis of TD96 is valid with this 
reinterpretation.

I want to emphasize that crustquakes in magnetars 
cannot be ruled out.\footnote{There is evidence for quite 
localized crust shifts during some magnetar outbursts.  
The radiative area of the thermal afterglow of the
1998 August 29 event was only $\sim 1$ percent of the neutron star area
(Ibrahim et al.~2001) consistent with an ``aftershock" adjustment along
a fault zone that was active in the August 27 flare. 
A similarly small radiative area was found following the 2001 April 28
burst which seemed to be an aftershock of the April 18 flare (Lenters et 
al.~2003; Feroci et al.~2003).
Observations of AXP 1E2259+586 during its June 2002 activity (Woods et 
al.~2003) showed an initial, hard-spectrum, declining X-ray transient with 
a very small emitting area during the first day following the glitch, while 
the  
emitting area of the slowly-declining thermal afterglow 
observed over the ensuing year
was a sizable fraction 
of the star's surface.  This suggests that there was a small region of the crust
where the magnetic field was strongly sheared, perhaps along a fault;
 and a large area in which it
experienced more distributed plastic failure.} 
Besides sudden yields of nuclear pasta and 
``mock fractures" involving fault-line liquification (\S 2), 
magnetic-mechanical instabilities in the outer layers of magnetars,
where magnetic pressure is not insignificant compared to the material pressure, 
may be associated
with the emergence of magnetic flux, as in solar activity
(e.g., Solanski et al.~2003). Wherever crustal fields exceed 
$\sim 10^{16}$ Gauss,
intrinsic magnetization instabilities may be possible (Kondratyev 2002).
Finally, rapid stress-changes exerted on the crust by 
the evolving core field from below, or by a flaring corona from above, 
could drive catastrophic failure.

\section{Conclusions}

In conclusion, a magnetar is a sun with a crust. Both crustal and 
coronal instabilities are possible, as well as 
instabilities within the core, which is coupled 
to the crust from below by the diffusing magnetic field. 
Physical conditions are much more 
complicated than those which prevail on either the Earth or the Sun, and 
the available data is much more fragmentary, so the challenge of 
understanding these stars is great.

In this review, I have described evidence that rapid interior stellar 
adjustments occur during some magnetar outbursts and bursting episodes. Based 
on this evidence, it seems likely that plastic crust failure 
initiates bursting episodes in SGRs and AXPs, by triggering a sequence of 
reconnection instabilities in the magnetosphere which are observed as 
``ordinary" common SGR (and AXP) bursts. Ongoing, relatively rapid 
plastic motion of patches of crust during these burst-active episodes 
(compared to what occurs in the quiescent state) could explain why 
bursts come in ``swarms" with the time between bursts much longer than 
the durations of the bursts themselves.

The ``relaxation system" behavior found by Palmer (1999) may be due 
to the (quasi)steady loading of magnetic free energy within the 
magnetosphere by the plastic 
motion of magnetic footpoints. Palmer's ``energy reservoir" would then 
be the sheared 
or twisted (i.e., non-potential) 
components of the exterior magnetic field, steadily driven by plastic
motion of the magnetic footpoints during active periods, and undergoing 
sporadic, catastrophic dissipation in bursts of 
reconnection.\footnote{This differs from previous suggested explanations 
of the Palmer 
Effect, which involved energy reservoirs within the crust.}
 If the reservoir is an arch of field lines with one footpoint anchored on
a circular cap of crust of radius $a$ that is slowly twisting at 
rate $\dot{\phi}$, 
then the loading rate is \hbox{$\dot{E}\sim (1/4) a^3 \,
B^2 \, \dot{\phi}$}, independent of the length of the arch in a first, crude
estimate. For a cap diameter $\sim 1$ km, comparable to the crust depth, this
implies $\dot{\phi} \sim 0.06 \, (B_{14}/3)^{-2} \ (a/0.5$ km$)^{-3}$ 
radians/day
during Palmer's interval B, and 25 times slower during Palmer's interval A. 
The durations of these burst-active intervals would then be time-scales for 
significant local crust-adjustments, exceeding by $\gtrsim 10^6$ the
time-scales for reconnection within individual bursts. 

There is little doubt that profound exterior reconnection occurs in magnetar 
flares (TD95).  There are two triggering possibilities: 

 {\bf (1)} A catastrophic, twisting crust-failure 
might occur during the flare, so that significant (magnetic) energy from
within the star contributes to the flare emissions.
If the solid crust yields plastically, then this would require a sudden
stress-change applied upon the solid from below; 
but crustal instabilities cannot be ruled
out (\S 5).
 
{\bf (2)} Flares might develop in the magnetosphere, with little energy 
communicated from below on the time-scale of the flare.  This could be 
a spontaneous instability reached 
via incremental motion of the magnetic footpoints; but ongoing
plastic failure
of the crust seems more likely as a trigger.

Clearly, these are not fully-distinct possibilities. Let us set the 
dividing line at 
$\sim 0.1$ of the energy coming from within.  Then, at present, 
I favor mechanism
(1) for the 1998 June 18 flare from SGR 1627-41; and mechanism (2) for the
1998 August 27 flare from SGR 1900+14.

The back-reaction of an exterior magnetic stress-change on the crust 
could drive shallow crust failure and heating that is consistent
with the August 27 flare afterglow (Lyubarsky, Eichler \& Thompson 2002).
But this back-reaction could not 
account for the 1998 June 18 afterglow
if interpreted as deep crust-heating (Kouveliotou et al.~2003).
The slow-peaking, tail-free June 18 event (Mazets et al.~1999b) 
plausibly involved a 
deep crust and/or core adjustment
in a star with a relatively quiet, relaxed 
magnetosphere, far from the critical state, 
so that little exterior reconnection 
was induced (relative to the giant flares).

Note that even the August 27 event 
was probably not a spontaneous, 
pure magnetospheric instability. 
A soft-spectrum precursor-event detected 0.45 s before the onset
of the 1998 August 27 event (Hurley et al.~1999a; Mazets et al.~1999a)
suggests that the crust was experiencing an episode
of accelerated plastic failure.
Plastic creep probably continued 
during the first $\sim 40$ seconds after the flare's hard spike, 
giving rise to 
the ``smooth tail" part of the light curve
(Feroci et al.~2001; \S 7 in TD01). Subsequent spindown measurements (\S 4)
suggest that large-scale currents in the magnetosphere
were enhanced rather than dissipated during 
the magnetically-active, flaring episode. 

\smallskip
Of course, much more work is needed to develop and 
test these hypotheses. 
Many mysteries persist, but it seems
that the magnetar model has
the physical richness needed to accommodate diverse 
observations of SGRs and AXPs.  The path to full 
scientific understanding
of these objects will no doubt be long and interesting. 

\smallskip
{\bf Acknowledgements}  \ I thank Chris Thompson, 
Malvin Ruderman, Ethan Vishniac and Peter Woods for discussions.
  
\smallskip
{\bf Publication Note} \ This paper will appear in 
{\it 3-D Signatures in Stellar Explosions,} 
eds.~P. Hoeflich, P. Kumar \& J.C. Wheeler (Cambridge 
University Press, 2004). 
The book version is somewhat abbreviated due to page limits.

\end{document}